# Semi-Transparent Image Sensors for Eye-Tracking Applications

*Gabriel Mercier[1][†], Emre O. Polat[1][†], Shengtai Shi[1][†], Shuchi Gupta[1,2], Gerasimos Konstantatos[1,3], Stijn Goossens[1,2]\*, Frank H. L. Koppens[1,3]\**

[1] ICFO - Institut de Ciències Fotòniques, The Barcelona Institute of Science and Technology, Castelldefels (Barcelona), Spain

[2] Qurv Technologies S.L., Castelldefels (Barcelona), Spain

[3] ICREA-Institució Catalana de Recerca i Estudis Avançats, Barcelona, Spain

[†]Authors having given equal contribution

\*Corresponding Authors :

Stijn Goossens : stijn.goossens@qurv.tech

Frank Koppens: frank.koppens@icfo.eu

# Abstract

Image sensors hold a pivotal role in society due to their ability to capture vast amounts of information. Traditionally, image sensors are opaque due to light absorption in both the pixels and the read-out electronics that are stacked on top of each other. Making image sensors visibly transparent would have a far-reaching impact in numerous areas such as human-computer interfaces, smart displays, and both augmented and virtual reality. In this paper, we present the development and analysis of the first semi-transparent image sensor and its applicability as an eye-tracking device. The device consists of an 8x8 array of semi-transparent photodetectors and electrodes disposed on a fully transparent substrate. Each pixel of the array has a size of 60 x 140 µm and an optical transparency of 85-95%. Pixels have a high sensitivity, with more than 90% of them showing a noise equivalent irradiance $< 10^{-4}$ W/m$^2$ for wavelengths of 637 nm. As the semi-transparent photodetectors have a large amount of built-in gain, the opaque read-out electronics can be placed far away from the detector array to ensure maximum transparency and fill factor. Indeed, the operation and appearance of transparent image sensors present a fundamental shift in how we think about cameras and imaging, as these devices can be concealed in plain sight.

Photodetectors, by being capable of converting light into electrical signals, have enabled a host of important technologies in the information era, due to their capability of translating visual information into digital data. These technologies are widespread, with applications in most industries, including smartphones, surveillance, automotive sensing, robotic systems, sorting, barcode readers, optical communication, medical imaging, movement detectors and night vision.[1] As a result image sensors, which are an array of photodetectors used to capture an image, are expected to reach a market size of $26.41 billion by 2024: a testament to their utility and our reliance on this technology.[2]

One particular application where the impact of image sensors has grown significantly is eye-tracking, as they offer a less invasive experience compared to specialised contact lenses, while providing better accuracy than electronic potential measurements.[3] Eye-tracking has a wide range of uses, such as detecting schizophrenia,[4] measuring the comprehension of texts[5] or driving experience,[6] while allowing for a better understanding of memory[7] and commercial choices.[8] Crucially, eye-tracking provides a human-computer interface[9] that can allow for the touch and gestureless control of automotive infotainment systems[10] – and is also earmarked as a key enabling technology for omnipresent virtual and augmented reality, by allowing foveated rendering to reduce display power consumption.[11] Significant efforts are being made to bring eye-tracking away from laboratory implementations, which tend to be bulky and require headrests and bite bars, to applications which allow for more natural head and body movements. The view of the user should not be obstructed, and discomfort should be minimised to allow for extended use of the eye-tracking device. Meeting these prerequisites opens up a host of possibilities in autonomous driving, both augmented and virtual reality (AR/VR), safety and security, 3D camera structures and environmental sensing.

Bringing eye-tracking to the consumer market is being explored and integration of eye-tracking directly into the glasses would omit any external instrumentation. So far, most designs of eye-tracking glasses are visibly different from traditional glasses, as they use one or several cameras mounted at an angle to properly extract pupil position. Such a structure generally has high computational complexity and thus power consumption, limits gaze accuracy due to the low resolution and angled position of the cameras used, occupies valuable space, and are also too conspicuous to be suitable for continual consumer use.

Utilising a semi-transparent photodetector array solves these issues as it directly integrates the camera onto the glasses, thereby reducing computational complexity while increasing portability through lower power consumption and an improved form factor. This method also discreetly integrates eye-tracking into everyday life, unlocking the possibility of widespread use. An existing approach towards transparent imaging involves patterning very small organic photodetector (OPD) pixels on transparent substrates.[12] This approach, however, suffers from small fill factors and low SNR.[13] Another semi-transparent photodetector fabrication method is to build OPDs with photoactive layers that are visibly semi-transparent while absorbing in the infrared,[14] however, this approach requires electronic read-out circuitry to be very close to the photodetectors, thereby reducing the transparency and fill-factor significantly when building an image sensor.

We propose a solution based on semi-transparent pixels with a built-in gain leading to a scalable semi-transparent image sensor with a high fill factor that does not compromise on transparency or portability. In this report, we show the image sensor's capability to capture still images as well as videos with a moving target that represents the pupil, demonstrating eye-tracking application. For eye-tracking applications, this sensor has a great deal of potential by offering the possibility to be placed in front of the users' eyes. This format allows seamlessly integrated portable eye-tracking as the sensor can be fabricated in the form of spectacles (Fig. 1c). The computing complexity of the eye-tracking is highly reduced as the full resolution of the captured image is centred on the eye. In addition to the natural semi-transparency of our detectors, an important innovation of our image sensor is the use of built-in photoconductive gain, which allows for separation between the semi-transparent photodetection array and non-transparent electronic read-out components. This overcomes the key challenge for conventional photodiodes that require read-out electronics in very close proximity to the detectors, which consequently reduces the fill-factor and prevents transparency. Moreover, our prototype 8x8 array is fully interconnected to further reduce the readout from 64 (8x8) to 16 (8+8) pins, offering a fully integrated system that is scalable to larger pixel density and array size. Finally, the detectors also have the capability to work in the SWIR regime with an eye-safe wavelength.[15]

**RESULTS AND DISCUSSION**

In order to provide an image sensor that is semi-transparent and exhibits high responsivity for the visible and SWIR range,[16] we use graphene/quantum-dot (GQD) pixels. Graphene, a two-dimensional crystal of carbon atoms,[17] provides both very high room-temperature electronic mobility (>100.000cm$^2$/Vs),[18] as well as transparency due to its atomic thickness.[19] Previous work has shown that graphene-based photodetectors can be applied in many application areas such as imaging[20] and data communications,[21] and offer advantages such as of flexibility,[22] CMOS integration,[23,24] broadband sensing,[25–29] and photo response speed.[30,31] Sensitizing graphene with quantum dots transforms it into a photoconductive system with responsivity reaching up to $10^7$ A.W$^{-1}$, while maintaining speeds of ≈1ms[32] and with combined sensitivity to both visible and SWIR wavelengths.[16] Due to the high mobility of the graphene, the devices exhibit built-in photoconductive gain. As a result, the photo-signals do not need amplification close to the detector. Instead, the signals can be transported to the edge of the array, allowing for the use of a non-local readout system placed away from the image sensor. Additionally, this technology allows for high fill factors: a little over 95% for QVGA (320x240 pixels) resolution (see SI for derivation of this value).

Our 8x8 photodetector array was processed on a transparent quartz substrate with pixels of 60 x 140 µm. The photodetector array was shown to have high transparency with respect to the bridge structures shown in Fig. 2b (85-95%, see SI Fig. 9). To allow for a non-local read-out scheme and high fill-factor, the pixels are interconnected by ITO contacts with elevated columns forming 3D bridge-like structures. This structure is depicted in Fig. 1a, which is a render of one part of the image sensor device. The graphene is then transferred and sensitised with a lead sulphide quantum dot layer, forming a photoconductive detector with photoconductive gain[16]. The bridge-like 3D ITO architecture yields the semi-transparent structure as seen placed in front of a pupil in Fig. 1b. The interconnectivity of the 3D structure allows for a standard multiplexed row-by-row readout, effectively reducing the amount of connections from the optical chip to read-out electronics from 64 (8x8) to 16 (8+8) wire bonds. The non-local readouts allow for non-transparent electronics to be placed away from the user's field of view.

As crosstalk between devices in the resistive array can be problematic, the readout electronics use a zero-potential scanning circuit, which allows row-by-row readout while avoiding crosstalk.[33] To compensate for device resistance variations, a dark voltage subtraction element has been added to the circuit. To avoid crosstalk between the pixels, we ensured that the resistance of the pixels is much higher than the resistance of the ITO leads. The basic electronic circuit diagram of a single powered row is shown in the right inset of Fig. 1c. In this circuit, the light-dependent hexagonal symbol represent the GQD pixels and the bottom op-amps represent difference amplifiers. To demonstrate the functionality of this design as a portable system, the electrical circuit is battery powered, which ensures the 50 Hz noise from the mains is avoided. This system reads a differential voltage on the circuit in order to keep the device isolated from the mains power line. Future development of an application specific integrated circuit (ASIC) with the read-out electronics would enable integration into the frames of glasses, as shown in Fig. 1c. The left inset represents the transparent GQD array structure.

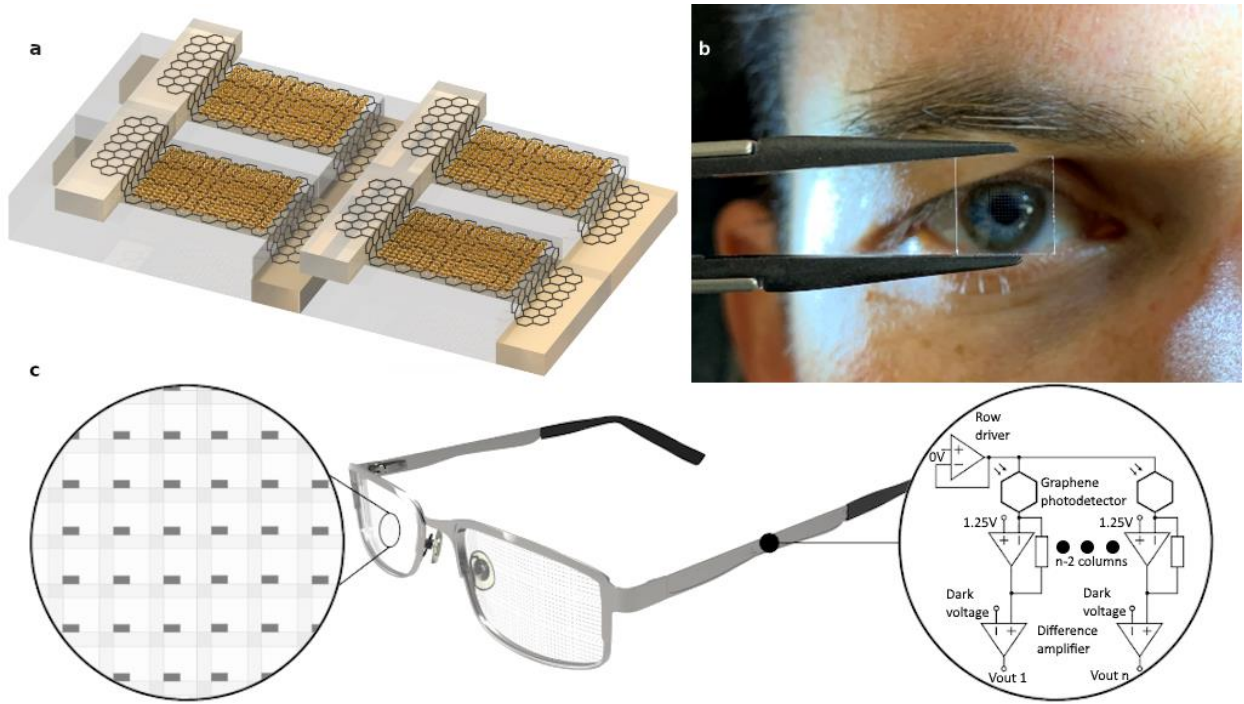

**Fig. 1 | Transparent image sensor concept. a**, Digital render of a 2x2 image sensor pattern, showing ITO contact lines (transparent gold), graphene structures (black hexagons), and quantum dots (copper dots). **b**, 8x8 array on quartz substrate and with ITO contact lines, displayed over an eye. **c**, Concept of the integration of the transparent image sensor and non-local readout electronics on spectacles. Left inset: schematic representation of the array structure. Dark grey pattern represents photosensitive area, while the light grey pattern represents ITO lines. Right inset: readout-electronic design used for each of the 16 contact lines.

To test the operation of our image sensor as a semi-transparent camera, we projected arbitrary shapes on the array, with a setup as shown in Fig. 2a. The patterns were actively illuminated with modulated light to minimise the electronic 1/f noise present in graphene.[34] The individual signals of each pixel were then extracted and normalised to their respective photoresponsivity, in order to reconstruct the image. The snapshot capabilities of our device are shown for a broad range of shapes, in comparison with the ground truth patterns, in Fig. 2b. In this figure, it is clearly visible that most patterns can be reconstructed by the imaging device. There is, however, occasional spreading of the shadows due to less than perfect alignment of the images on the array (as visible in Fig. 4).

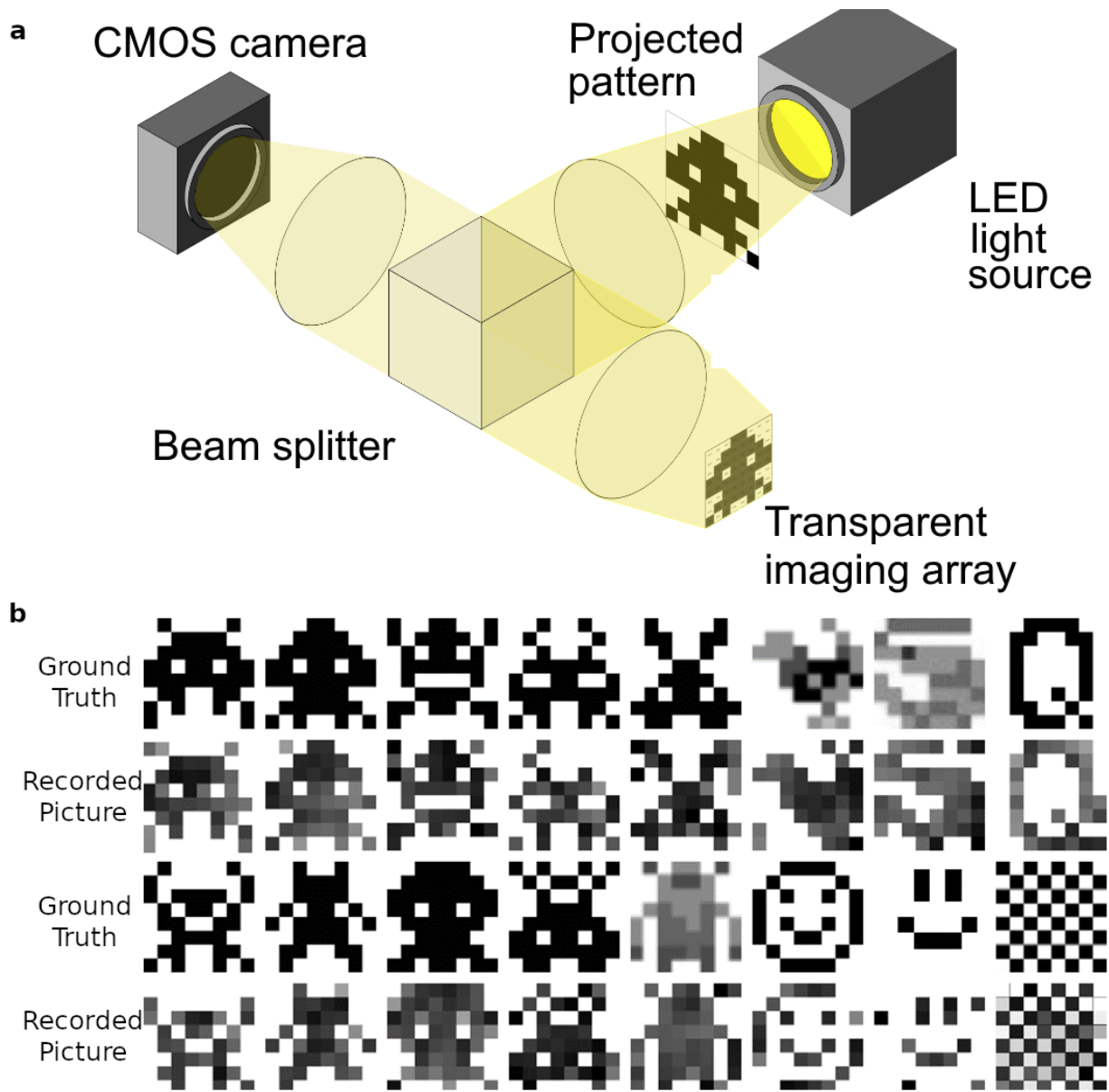

**Fig. 2 | Transparent image sensor results**. **a,** Experimental setup diagram. A CMOS camera allows for real-time monitoring of the projected pattern onto the transparent sensor. **b,** Ground truth projected patterns alongside the recorded images from the transparent image sensor. The LED light source modulated at a frequency of 500Hz. The recorded voltage amplitude for each pixel at the modulation frequency was displayed as an image. The images are greyscale pictures, where the lightest value was calibrated with a fully illuminated device, and the darkest value calibrated when the device is in full darkness.

## Device Performance

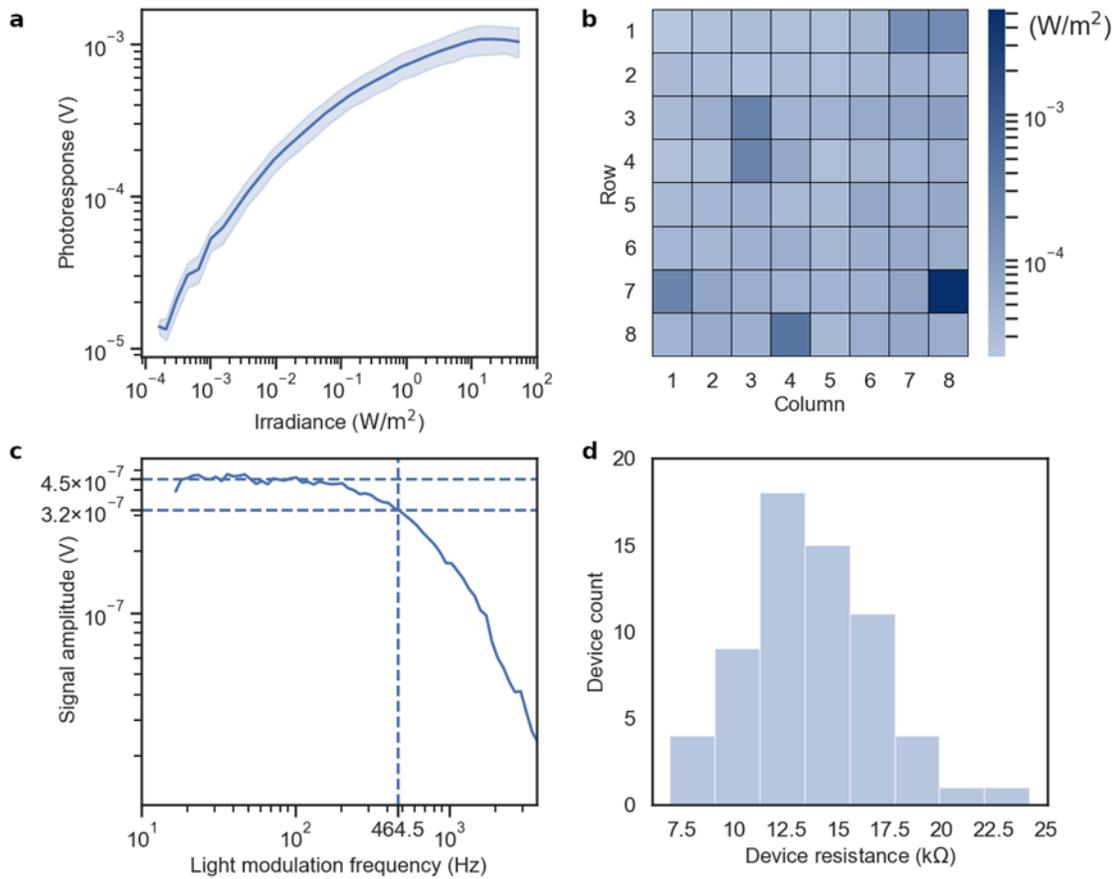

**Fig. 3 | Opto-electronic performance of the transparent image sensor. a,** Mean power dependence of the array photoresponse, plotted with the 95% confidence interval as a blue shadow. This measurement was performed using a 637 nm laser light source, with 1 Hz modulation. **b,** NEI map showing devices with relatively low NEI (<$10^{-4}$ W.m$^{-2}$ for wavelengths of 637 nm). **c,** Photodetector sensitivity as a function of LED modulation frequency with power output of 23.1 mW. **d,** Resistance histogram of 63 photodetectors of the image sensor. One pixel of resistance of 100 kΩ was omitted.

A key metric in assessing the quality of the detector array is a uniformity evaluation of the photosensitivity. The uniformity of the pixels is displayed in Fig. 3a, which shows the power dependence of all pixels in the array. In this figure, the line represents the mean, while the 95% confidence interval is shown as the blue area around the mean. As there is little spread around the mean value, we can conclude that the pixels are working with a high level of uniformity. The power dependence shows a full dynamic range of at least 45dB, where the noise-equivalent-irradiance (NEI) is limited by the 1/f noise of the pixels. Fig. 3b shows the noise equivalent irradiance map of our 8x8 pixel array. Even though there is an uneven distribution, most likely due to the initial graphene quality variation, all the pixels are photosensitive and more than 90% of the pixels show very high sensitivity (NEI < $10^{-4}$ W/m$^2$ for wavelengths of 637 nm, see SI for further information). A histogram of pixel resistances is shown in Fig. 3d. This figure shows a relatively constrained set of resistive values. The resistive 2D map (shown in the SI Fig. 7) shows the series resistance of the ITO (due to the not optimized ITO deposition recipe), with higher resistance values closer to row 8 and column 8 (as expected since the current has to travel on ITO lines for longer distances). Improvements of the ITO deposition recipe will remove the series resistance effects imposed by the lines.

A crucial characteristic for eye-trackers is the refresh rate, which should exceed 200Hz for effective operation.[35,36] We measured the maximum speed of our detectors by measuring the pixel response to a modulated LED, and by sweeping through a range of modulation frequencies. As seen in Fig. 3c, the cut-off frequency for the pixels is 465 Hz. We estimate that the frame rate of the full array is determined by this cut-off frequency (see SI). As sampling errors are not a large issue at this speed, we conclude that the speed of our photodetector arrays fulfil the requirements for eye-tracking measurements.

To demonstrate eye-tracking capabilities, the optical setup was modified to allow a moving black dot to be projected across the array. Additional software was written to enable real-time tracking of the dot. The software displays raw sensor data and allows us to manually set a threshold to clean up the image. Active light modulation of 723 Hz was used to suppress 1/f noise (as this is above the cut-off frequency, the image sensor has a reduced photoresponse of detectors, in exchange of higher frame rates). Fig. 4 shows four screenshots of the real-time monitoring of the array. The top row shows a live feed from a CMOS camera, while the bottom row shows the raw data from the image sensor. The live video runs at 13.5 fps (frames per second, limited by the read-out system), which allows us to monitor any sort of movement from the black dot (see SI for video footage). The current system allows monitoring of the movement of the black dot in real time, demonstrating proof of concept eye-tracking and other types of instantaneous monitoring applications. We remark that the framerate is currently limited by the generic data acquisition system used. Moving to a dedicated read-out circuit or ASIC would greatly improve the framerate. The theoretical limit of the frame rate for these arrays has been calculated at around 20,000 fps for an array with QVGA resolution, or 240 x 320 pixels (see SI).

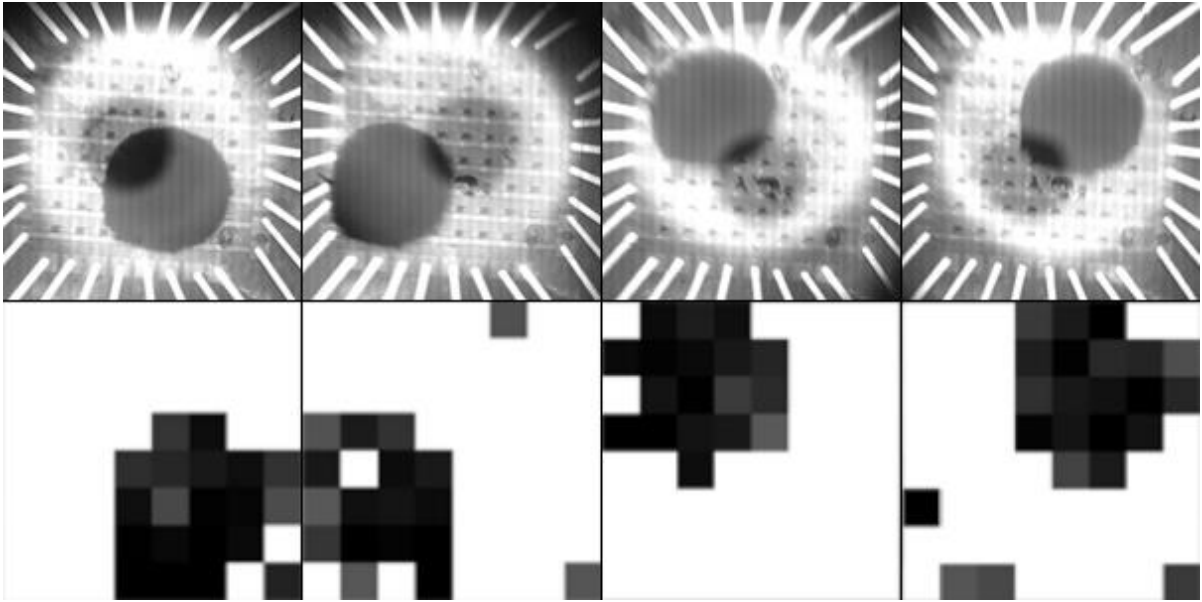

**Fig. 4 | Eye tracking simulation experiment.** The top panels show the projection of the black dot on the image sensor array, superimposed on a recorded image recorded by a CMOS image sensor. The four pairs of bottom panels show the image sensor signal output, with the dot placed at four different positions. These images are screenshots from a live video shown in the SI, in which the dot is tracked moving around the array. The images have been thresholded in real-time in order to only show significant dark patterns in the image. The dots have a 1.6mm radius and are illuminated by a 723 Hz modulated light source. (see SI for video footage)

### Conclusion

Future transparent image sensors based on the presented approach can be scaled to much larger pixel counts because the interconnected m x n pixel structure is scalable. With a theoretical fill-factor higher than 93%, this image sensor technology is ideal for a fully transparent, high-definition camera with a high fill factor, capable of real-time measurements. The infrared sensitivity allows for the use of an active light source undetectable to the human eye, ensuring both low-noise and lens-free operation, for example by employing smart focusing strategies such as visibly transparent IR pinholes,[37] or Fabry-Perot-style structures in order to forgo coloured filters.[38]

Our approach to eye-tracking implementation has the potential to accelerate the adoption of eye-tracking enabled devices in fields ranging from AR/VR to automotive. As the pixels of the array are both semi-transparent and flexible,[22] eye-tracking could be implemented in previously unimaginable form factors, such as on prescription glasses or curved windshields. Adding image sensing capabilities to glass-like materials opens up a wide range of application previously unexplored within the realm of image sensors such as adding image sensor capabilities to building windows or transparent screens. Additionally, as these image sensors allow for 3D stacking[39], more complex architectures that can perform light field measurements[40], phase searching or transparent and lens-less behaviour can be developed.

**Methods**

We implemented a lithography process with photoresist (AZ 5214) to form necessary patterns, then sputtered 100nm of indium tin oxide (ITO) on the quartz wafer, and finally, desired rows structures are obtained after a lift-off process. The ITO is then annealed at 300 degrees Celsius in an ambient environment for approximately 5 minutes until the transparency of the ITO is increased visibly and yielding a sheet resistance of ≈51 Ohm/□. We then use atomic layer deposition to deposit a layer of aluminium oxide, (AlOx) with a thickness of 50nm, as a dielectric to form the bridge structure between rows and columns. After another lithography process, numerous contact windows are opened on the dielectric layer by reactive ion etching. Then, columns are obtained on top of the dielectric layer using the same methods previously described to make the rows, and the whole structure is once again annealed in the same fashion to increase the transparency of ITO.

Small buffer layer strips of titanium palladium with a thickness of 30nm are deposited by e-beam at the previously opened contact windows in order to form a good contact with graphene subsequently, to avoid direct graphene-ITO contact which gives very high resistance values. The graphene is then transferred using a wet transfer method, patterned using lithography and reactive ion etching, and in the end sensitised with a lead sulphide quantum dot layer, using a simple spin-coating step. In order to synthesise the quantum dots, lead oxide (PbO) (0.45 g), oleic acid (1.5 mL) and 1-octadecene (ODE) (18 mL) were placed in a round bottom flask and pumped under vacuum overnight at 95 °C. After 12 hours, the flask was shifted to Argon and its temperature was increased to 100 °C. In a separate vial, under nitrogen atmosphere, 210 µL of hexamethyldisilathiane (TMS) was mixed with 10 mL of pre-dried octadecene. This solution was injected into the Pb-oleate mixture at 100 °C. Subsequently, the heating of the reaction was switched off and the flask was allowed to cool down to room temperature. The PbS QDs were cleaned and extracted from the reaction mixture by adding acetone via centrifugation. The process was repeated 3 times and finally the dots were dissolved in toluene for spin coating.

**Acknowledgements**

The author would like to thank and acknowledge the contribution of Marc Montagut in providing 3D structural renders, Silvana Palacios for providing invaluable help in setting up the optics, Tymofiy Khodkov for performing AFM and Raman on samples as well as helping with the ITO lift-off, Julien Schreier for providing guidance in estimating the maximum theoretical framerate of the devices, the ICFO electronic workshop for helping in the realisation of the readout circuit, Johann Osmond for ITO characterisation, in addition to Pablo Oscar Vaccaro who not only helped with ITO characterisation but creating the current recipe used for ITO deposition.

F.H.L.K. acknowledges support from the Government of Spain (FIS2017-91599-EXP; Severo Ochoa CEX2019-000910-S), Fundació Cellex, Fundació Mir-Puig, and Generalitat de Catalunya (CERCA, AGAUR, SGR 1656). Furthermore, the research leading to these results has received funding from the European Union's Horizon 2020 under grant agreements no.785219 (Graphene flagship Core2) and no. 881603 (Graphene flagship Core3), and the ERC proof-of-concept grant 786285 (Gtrack).

## Author Contribution

G.M., E.O.P., S.Go and F.H.L.K. conceived and designed the experiment. G.M performed measurements and data analysis and wrote the manuscript. S.S. wrote the device fabrication section of the manuscript. E.O.P and S. S. fabricated the samples and performed characterizations. S.Gu synthesized colloidal quantum dots. E.O.P., S.Go and F.H.L.K. supervised the work. All authors contributed to the manuscript revisions. All authors contributed to the scientific discussion.

## Competing Interests statement

The author states no known conflict of interest with the presented work and research.

# Supporting information

- Optical setup
- Picture extraction
- Extraction of the photoresponse
- Noise Equivalent Irradiance (NEI) map
- Extraction of external quantum efficiency
- Fill factor calculation
- Resistive heat map
- Hardware design
- Transparency

# Supporting videos

Dot.mov : representation of the movement of the eye iris

Checkerboard.mov : demonstrating the imaging capability using the blinking of the pixels in a checkerboard pattern

# Supplementary Information

## Optical setup

The optical setup, a simplified representation of which is shown in figure 2a) of the paper, was built using Thorlabs cage mounts. The cage mount allowed changing between two light sources:

- Thorlabs MWWHF2 LED of temperature color 4000K (used for creating the projections and measuring the speed of the devices)
- Thorlabs LP637-SF70 is a laser of wavelength 637nm which was used to measure the power dependence curve of Figure 3a), and the NEI map of figure 3b)

The projected patterns were printed on acetate in black and white, and were wedged between two Thorlabs WG11010 optical windows within the cage system. The setup then used a Thorlabs LA1608 Plano-Convex lens with a focal length of 75 mm, to put the pattern in infinity focus. The pattern is reflected off a Thorlabs PF10-03-F01 aluminium mirror placed on a Thorlabs KCB1C/M right angle kinematic mount, allowing us to perform large movements with the projected pattern.

A beam splitter is present to allow a DCC1545M black and white camera to observe the projected pattern on the transparent photodetector. A Thorlabs LA1608-B plano-convex lens of focal length 75mm is used in order to focus the pattern on the transparent photodetector, while a Thorlabs LB1901-B-ML biconvex lens of focal length 75mm is focuses the light coming from the photodetector into the camera. This optical setup allows fundamental measurements on the image sensor, as well as the capability to record videos on both the image sensor and CMOS camera simultaneously.

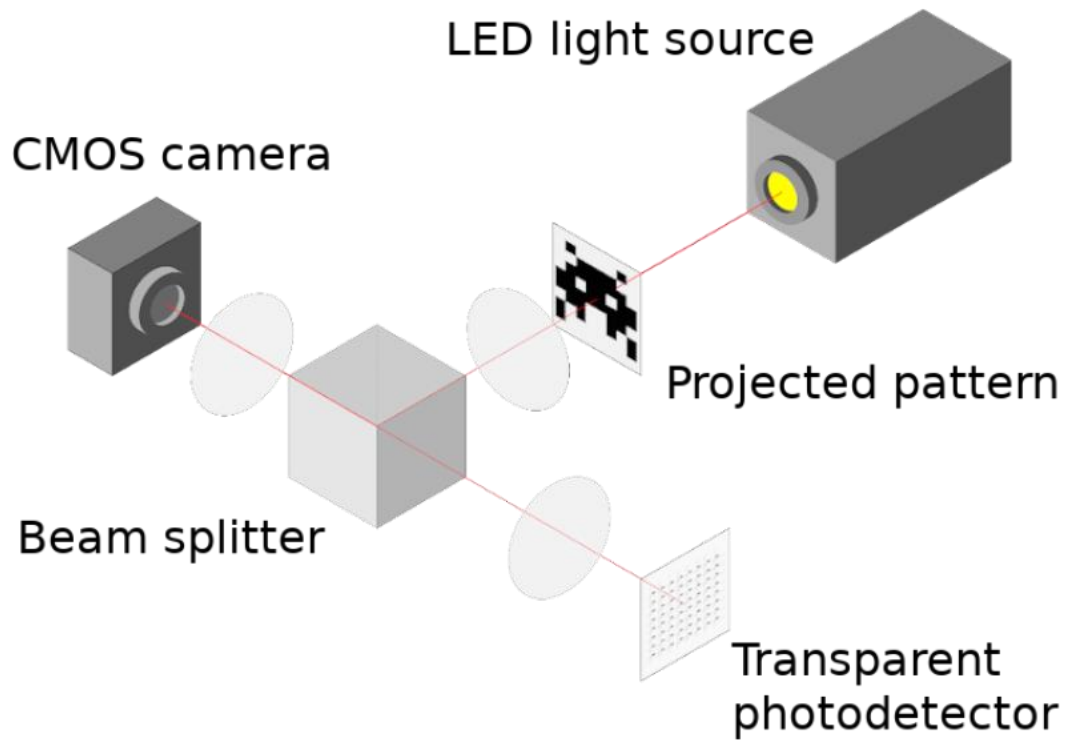

SI Figure 1: Optical setup schematic

## Picture extraction

Extracted pictures are present in figure 2b) of the paper. Prior to taking pictures, the device signal ranges are calibrated. In order to do this, we measure the response of the array when subjected to a dark environ, and subsequently when subjected to the LED light source. This allows us to linearly map our pixel photo response from fully illuminated to receiving no light. The duration of the calibration measurements ranged from 10 to 30 times the period of the modulation of the LED light, although this timing can realistically be shortened. The signals were samples at 125 kHz, but this signal can be reduced to the Nyquist rate (2 times the LED modulation frequency). The same measurements were then made when projecting a set of figures. The projections were done by printing greyscale images on a transparent slide (acetate).

The data analysis uses standard libraries in python: numpy, scipy, pandas and openCV. The code first opens the calibration data from both previously mentioned measurements. They are then transformed to frequency domain via fast Fourier transform, and the amplitude at the modulation frequency is stored in two individual matrices (one for the maximum illumination, and one for no illumination). The captured data of the image goes through the same data processing steps and then is linearly mapped between the maximum and minimum illumination values.

Linear mapping does not account for the logarithmic behaviour seen on the device, and so the intensity of the final picture does not exactly match the intensity of the incident light. It is, however, a good approximation considering the reduction in computation complexity of this method. Non-linear mapping or a lookup table could be considered if such approximations were not acceptable. The mapped images of the pictures were put through the "Inverted to zero threshold" function in opencv in order to clean up small fluctuations in pixels which were fully illuminated.

## Extraction of the photoresponse

The photoresponse is shown in figure 3a) of the paper. The photoresponse values were taken in a two-step process. We use an optical setup with a digital variable attenuator (OZ Optics da100) to control the collimated output light. A power sensor is placed at the output of the light (Thorlabs S120C), which then records the light power for the full range of attenuation values of the digital variable attenuator. The irradiance is then extracted from the geometrical size of the power sensor. The image sensor was then placed instead of the power sensor, the light was modulated, and the photodetectors took measurements for the full range of attenuation values of the digital variable attenuator. The data was processed in python, where the fast Fourier transform of the signal was taken, saving the amplitude of the signal at modulation frequency for each attenuation value. The signal was then mapped against irradiance using the power sensor data.

## Noise Equivalent Irradiance (NEI) map

The NEI map is shown in figure 3b) of the paper. The noise floor of the system was calculated by taking the power spectral density of devices. We take the mean of the white noise (non-1/f noise). This value is taken for each device in the array and provides us with the noise floor of each device.

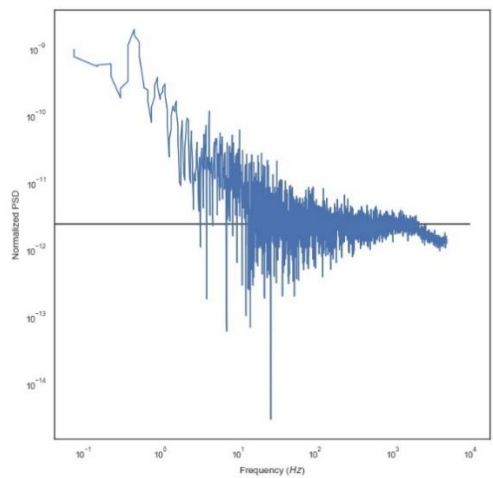

SI Figure 2: Noise floor extraction from power spectral density. The horizontal line represents the extracted noise value

The photoresponse curves of each device were then fitted with a curve of shape $ax+bx^{0.5}+c$ in the log-log scale. The intersection between the fit and the noise floor was then extracted in order to give a single NEI value per device:

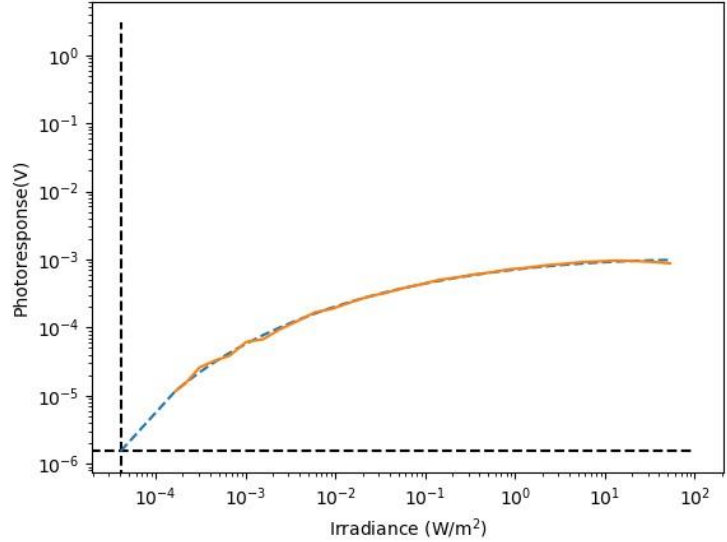

SI Figure 3: NEI extraction from intersecting photoresponse and irradiance. The measured photoresponse is shown in orange. The light blue curve shows our fit. The black horizontal line represents the device noise floor. The black vertical line represents the NEI.

## Extraction of external quantum efficiency

The external quantum efficiency has been calculated using the following equation which dictates the photo-resistive change in our device : $\frac{dR}{R} = \frac{Q.E \times \tau}{E_p \times n} \times Irr$

Where Q.E represents the quantum efficiency, τ the integration time of the detectors, Irr the irradiance on the detector, $E_p$ the photon energy, and n the charge carrier density.

We can extract the resistive change $\frac{dR}{R}$ from our measurements as 0.03, as well as τ (1.75 ms) using a defined irradiance (211.6 W/m²). Photon energy is defined by the wavelength of laser used (637 nm). The charge carrier density is extracted through the equation dictating the resistance of the devices:

$$R = \frac{1}{\sigma}\frac{L}{W}$$

Where R is the device resistance, σ the graphene conductivity, L and W the length and the width of the device respectively. The conductivity of graphene relies on its charge carrier density according to the following simplified equation:

$$\sigma \approx ne\mu$$

With n the charge carrier density, e the electron elementary charge, and µ the electron mobility

Thus by assuming a mobility of 1000 cm$^2$/V.s, and plugging the values into the equation we calculate a quantum efficiency of approximately 5.4%.

## Fill factor calculation

Fill factor was calculated based on the required resistance needed in photodetectors to offset the crosstalk effects caused by ITO. Assuming we scale the current architecture to QVGA resolution, the total resistance contribution of ITO has been calculated to be 126 016 Ω using a sheet resistance of 36 Ω/□.

To offset the voltage drop on the ITO lines, we want the resistance of the devices to represent 99% of the total resistance. Therefore, we want devices of resistance 12 475 584 Ω. The resistance of graphene primarily dictates the resistance value of our photodetectors. Assuming a resistivity of 2 kΩ/□, the pattern of our device needs to have 6238 squares of graphene.

A single element in the pattern of our array measures 350 * 350 µm. This means that the structure of a device is limited to this size. With our current laser writing capabilities, we are hardware limited by a resolution of 1µm. In order to avoid shorting all image sensors, we trim 1 µm around this array, allowing the device to be designed on a 348 * 348 µm area. The optimal strategy for building squares is by removing 1 µm line at 2 µm intervals, essentially allowing 348+1 squares to be designs on a 348 * 2 square area. If we round up, we can calculate that 6238 squares fit on 18 lines of size 349. Therefore, by removing 18 lines of length 347 µm, the design of our detectors fill the resistance requirements. By taking into account both the removal of the "frame" and the lines, we can calculate the fill factor as follows:

$$\frac{350 \times 350 - (17 \times 347 + 350 + 350 + 349 + 349)}{350 \times 350} = 0.9376$$

Giving us a fill factor of 93.76%.

## Max framerate

The current image sensors have a structure similar to the form shown in SI figure 4, where the dark grey rectangles represent the GQD detectors, and the light grey lines represent the ITO columns and rows. This pattern is repeated X number of times to form the image sensor. For QVGA resolutions, the pattern is repeated 80 times horizontally and 60 times vertically (to represent a 320x240pixel sensor).

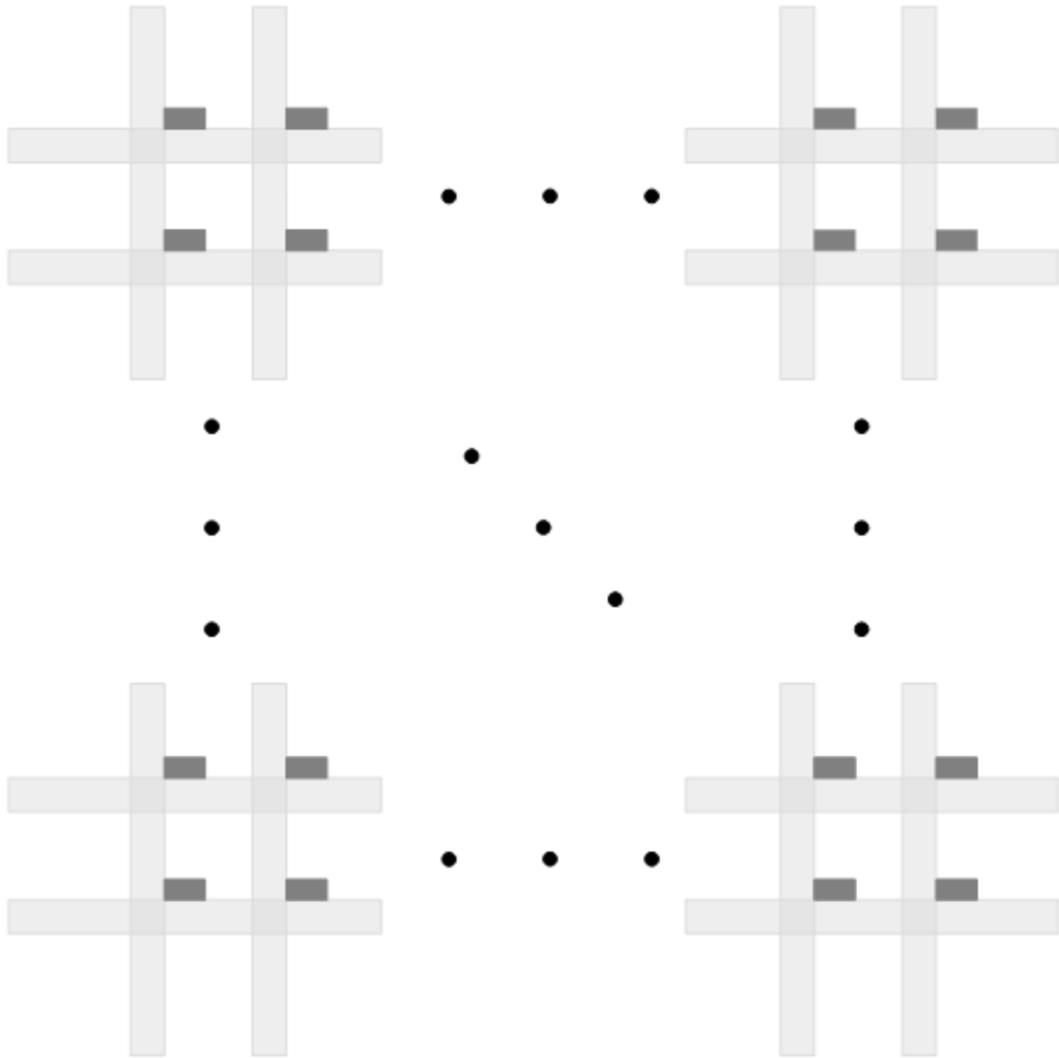

SI Figure 4: Current structure of the image sensor

Two parasitic capacitances have been identified in the current structure. One is present when bridging rows and columns of ITO, while another inherent capacitance is associated with the graphene devices.

These areas are highlighted in red in SI Figure 5.

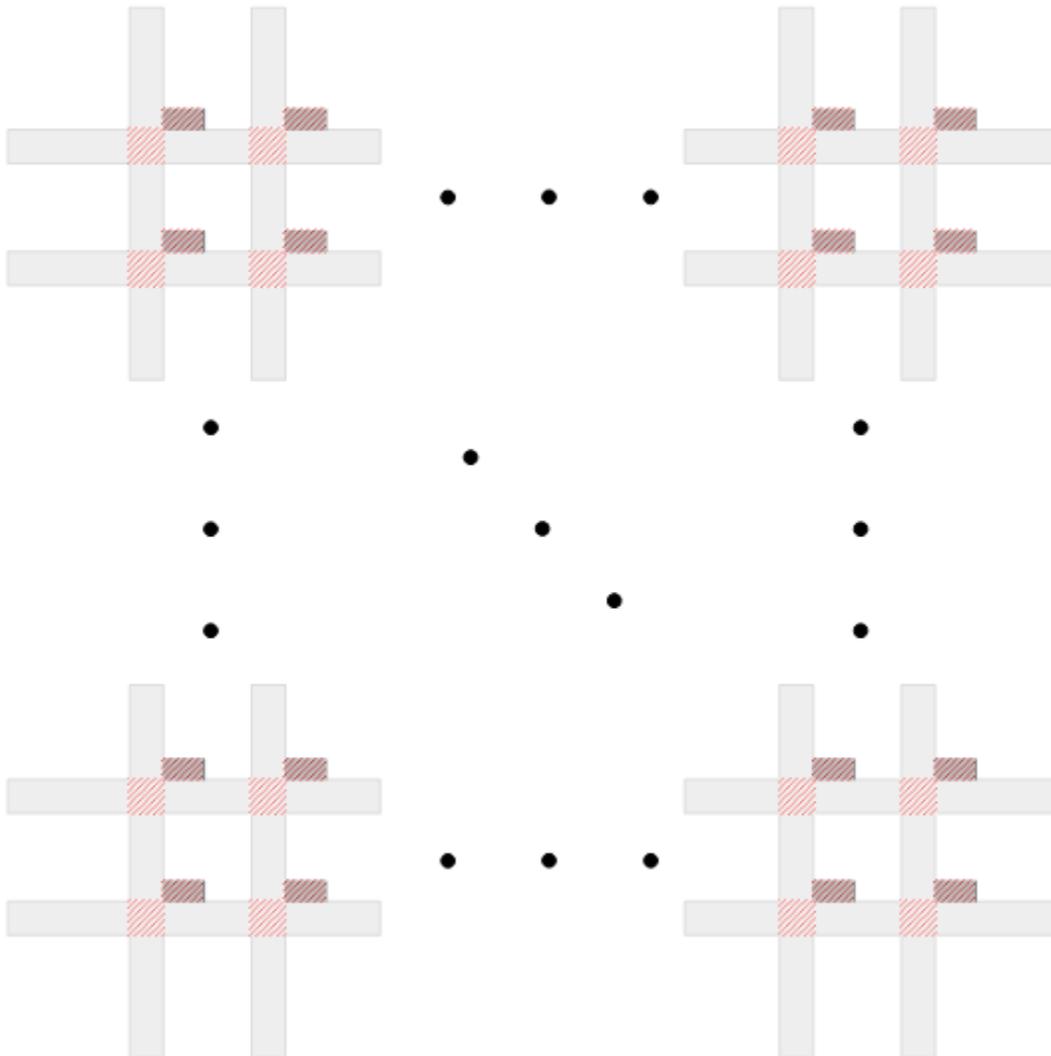

SI Figure 5: Possible parasitic capacitances in the image sensor

It is useful to note that in the way the electronics are currently operated, the readout is performed in the following fashion: A row is set to a drive voltage by an op-amp, which would be placed on the right hand side of figure 5. The other rows are left at a reference voltage potential. The electronics then read the values column-by-column with op-amps that would be present at the top of figure 5.

When a row is selected, the two capacitances could plausibly affect the signal.

We simulate the transient behaviour of turning on the first and the last row of a QVGA resolution device, while reading out the first and the last column of the sensor. These columns should show the most extreme behaviours present in the sensor. The resistance of the ITO was extracted using the resistivity of 36Ω/sq. The upscaling to QVGA resolution was done by replicating the current geometry of the device until 240 by 320 devices were present.

$$\text{The bridge capacitance was obtained using the equation: } C_1 = \frac{\varepsilon_r \varepsilon_0 A}{d}$$

With the current geometrical bridge structure, we get a capacitance of 81.2 fF. The capacitance inherently present in the devices was considered in the model which scales with the device size at the rate of 3 fF.µm$^{-2}$. As the size of the devices are 60 by 140 µm, we have parasitic device capacitance of 25.2 pF. The resistance of the devices was derived from their expected value, while considering their geometrical shape, giving us 1.3kOhms per pixel. Using these values, we modelled the following simulation circuit in LTspice, using ideal op-amp models:

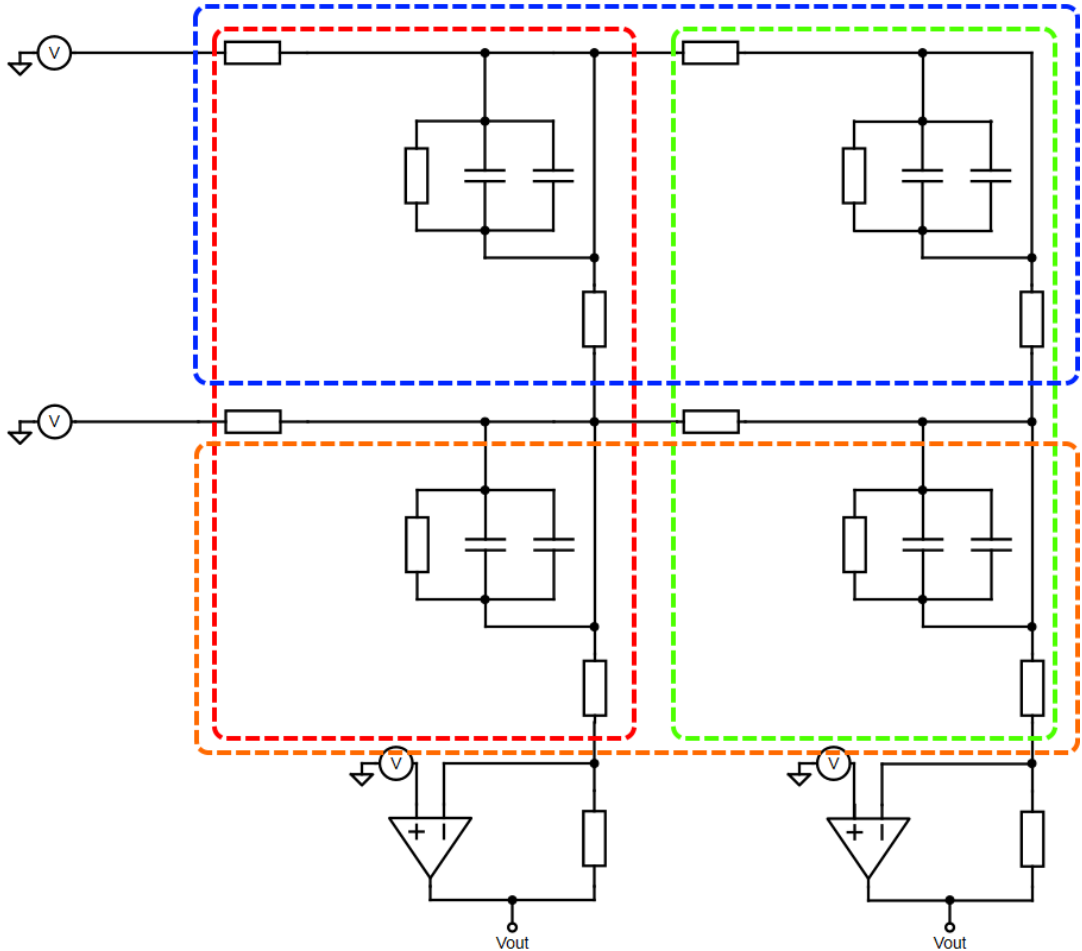

SI Figure 6: Schematic of model circuit for switching in LTspice

When powering a row, the signal of the last column is influenced by the capacitance in all the previous columns of this row. Thus, for LTspice, we alternate between putting the cumulative capacitive events in the red left hand side sub-circuit, and the capacitive effects of a single row in the green right-hand side sub circuit and vice versa. This simulates looking at the furthest and the closest column respectively.

We simulate the voltage output when switching from applying a driving voltage between the blue and the yellow half. In order to extract the transition values of the output, we calculate the time at which we reach 90% of the final output value. Using this method, we get the following results:

|  | Row 240 | Row 1 |
|---|---|---|
| **Column 320 off-time (s)** | 3.73443E-08 | 7.55E-08 |
| **Column 320 on-time (s)** | 3.73469E-08 | 7.54939E-08 |
| **Column 1 off-time (s)** | 4.06312E-08 | 3.32592E-08 |
| **Column 1 on-time (s)** | 4.05844E-08 | 3.32631E-08 |

Table 1: Switch times according to LTSpice simulations

We see the longest simulated transient takes approximately 75.5 ns. Assuming a worst case scenario where the voltage of a column would need to be switched off then on again, and that this transient timing would be applied to the entirety of the array, a conservative estimate shows that refresh rates of a QVGA camera could reach 1/(7.55263071499801e-08*2*320)≈20 kHz. We can thus conclude that the 465Hz cut-off frequency mentioned in the paper is due to the speed of the fabricated photodetectors and not their fundamental structure. As fabrication methods improve, we will come closer to the calculated 20kHz limit.

## Resistive heat map

Building a two dimensional heat map of the image sensor allows us to gather information on the quality of the fabrication process. Ideally, one would want a uniform quality of graphene for each pixel to behave similarly. However, we have experimentally found that graphene tends to display a variation in its properties even within a single sheet.

Within the image sensor, we expect two materials to be the major contributors in electronic resistance: the graphene of the photodetectors and the ITO interconnect. We expect the resistance of the ITO to increase linearly (due to the geometry of the image sensor) the further away we get from the first row and column. We also expect resistance changes due to graphene defects to be randomly present within the sheet of graphene. By building a heat map of the resistance of the image sensor, we get a visual sense of the degree in which we are affected by the random variation in graphene as well as the quality of the ITO deposition.

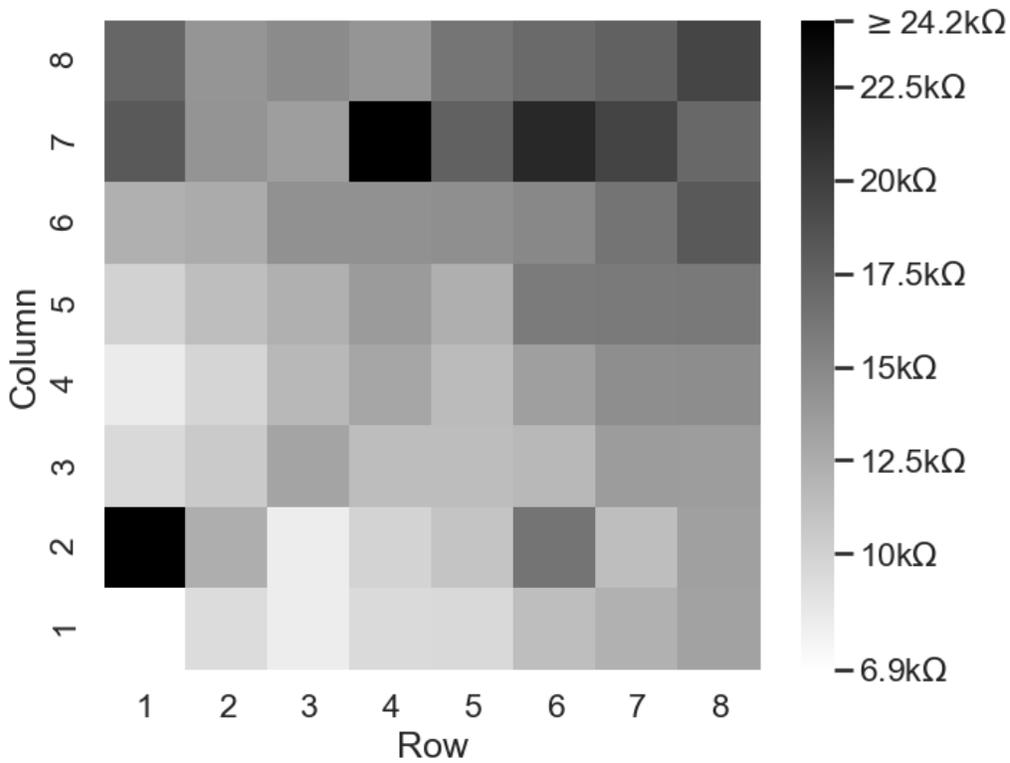

SI Figure 7: Resistance heat map of the image sensor

We measure the sheet resistance of our ITO to be ≤40 Ω/□. An initial visual inspection of the heat map reveals a gradient pattern starting from the bottom left corner, moving towards the top right corner. This gradient stemming from the ITO shows we have relatively little variation in resistance due to graphene defects, or when creating our ITO-graphene interface.

## Hardware design

The main consideration when designing the readout electronics was to provide a design that is portable, as the final product would need to be integrated into spectacles. This means that the readout electronics would be battery powered, which also allows us to construct a low noise measurement system that is free of contamination from the mains. The design of the electronic hardware was drafted with the idea of modularity at its core, as this provides a solid foundation, while still allowing testing of additional features at later stages.

We used a NI PXIe-6363 Multifunction I/O Module, housed in a NI PXIe-1073 chassis. The PXIe-6363 has 16 differential analogue inputs. Differential analogue inputs are crucial, as we want to measure voltage differences on our portable PCB board and not compare voltages with the mains ground. The 16 inputs give us the potential to scale our design to 16x16. The PXIe-6363 also has 48 digital inputs and outputs, which can work with speeds of up to 100 MHz (dictated by the internal clock of the chassis). We use the PXIe-6363 to read the output signal of our readout electronics, as well as digitally communicate with the components on the board.

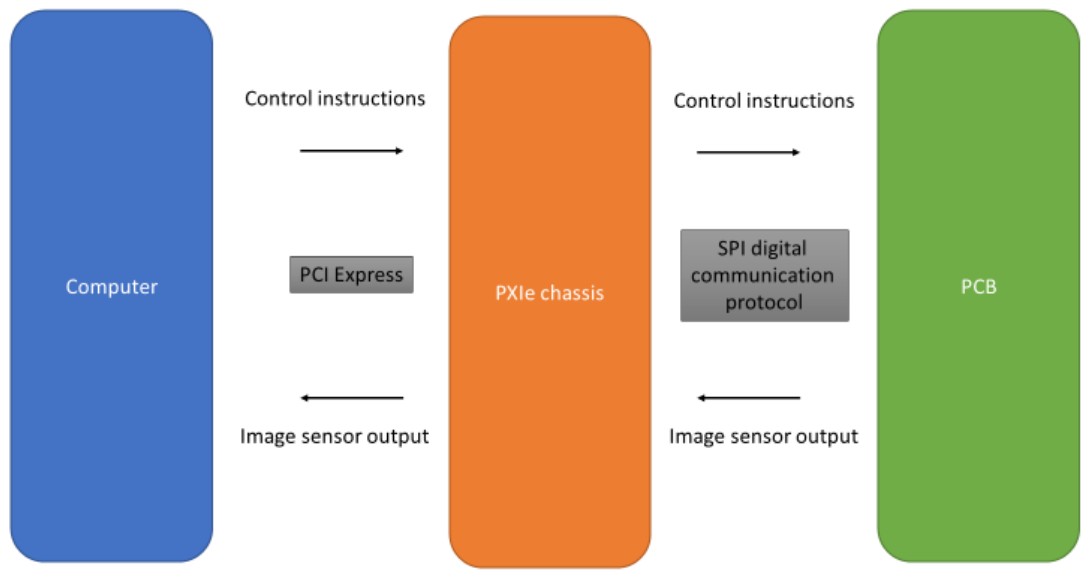

SI Figure 8: Block diagram of communication from computer to the electronic circuit

The PXIe-6363 needs to communicate with the multiplexers, which select which row to power, and the DACs, which set the dark voltage at PCB level. The instructions for the PXIe-6363 to communicate using a Serial Peripheral Interface (SPI) protocol. The low-level instructions were coded on the computer in order to translate higher level instructions to properly timed bits turning on and off at specific digital output of the PXIe.

This implementation has been successful in showing the ability to capture images as shown in the paper, but is slow as the computer needs to continuously translate all input instructions, then transfer these instructions to the PXIe. The PXIe does not have the memory required to allow it to save a set of instructions to follow. This means that the computer constantly needs to upload the instructions to power a new row and read sensor data, which is too costly if we want to reach very fast framerates. One way to reduce the size of the electronics as well as speed up the measurements is to use a microcontroller to replace the PXIe-6363 in its operation.

## Transparency

The transparency of the device was measured using a Cary 5000 spectrophotometer. For this measurement, we used the quartz substrate with ITO lines as a baseline to measure the transparency of pixels. This yields the following results:

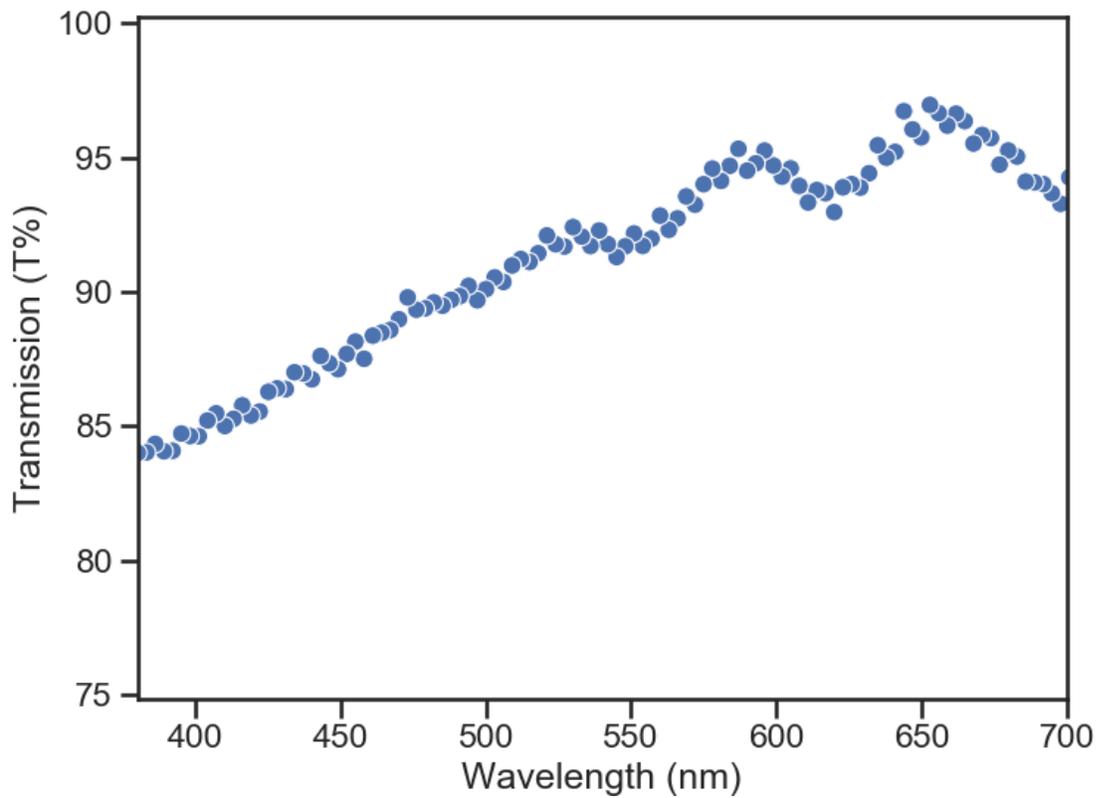

SI Figure 9: Transmission of the image sensor, using the ITO-quartz structure as a baseline

We calculate that, on average, each pixel allows 91.2% of visible light through. Literature shows graphene to absorb around 2% of the light, and thus we estimate quantum dots to absorb approximately 6-7% of the incoming visible light[19].